\documentclass[preprint,showkeys,nofootinbib]{revtex4}
\usepackage{hyperref}
\usepackage{epsfig}
\usepackage{graphicx}
\usepackage{amsmath,amssymb}
\usepackage{bm}
\usepackage{color}
\usepackage{empheq}
\usepackage{slashed}
\usepackage[all]{xy}

\newcommand{\beq}{\begin{equation}}
\newcommand{\eeq}{\end{equation}}
\newcommand{\bea}{\begin{eqnarray}}
\newcommand{\eea}{\end{eqnarray}}
\newcommand{\bean}{\begin{eqnarray*}}
\newcommand{\eean}{\end{eqnarray*}}

\newcommand{\wD}{{\widetilde D}}

\newcommand{\ws}{{{S}}}

\newcommand{\Ar}{{\tilde r}}
\newcommand{\Atheta}{{\tilde \vartheta}}
\newcommand{\Aphi}{{\tilde \varphi}}
\newcommand{\At}{{\tilde t}}
\newcommand{\Arho}{{\tilde \rho}}

\begin{document}

\title{The definition of mass in asymptotically de Sitter space-times}

\author{Brian P. Dolan}

\email{bdolan@thphys.nuim.ie}

\affiliation{Department of Theoretical Physics, Maynooth University\\
  Maynooth, Co.~Kildare,  W23 F2H6\\ Ireland\\}

\affiliation{School of Theoretical Physics\\
  Dublin Institute for Advanced Studies\\
  10 Burlington Rd., Dublin, D04 C932 \\Ireland}

\preprint{DIAS-STP-18-10}

\begin{abstract}

An invariant definition of mass in asymptotically de-Sitter space-times is given 
that relies on the existence of a time-like Killing vector on a sphere surrounding the mass but does not require going to an asymptotic region. 
In particular the mass can be calculated exactly on a sphere inside the cosmological horizon.  The formalism requires varying the background metric solution by a perturbation that satisfies the linearized equations of motion but need not share the Killing symmetry of the solution and is therefore ideally suited to calculating masses in stationary space-times perturbed by a gravitational wave without going beyond the cosmological horizon.

\end{abstract}



\maketitle

\section{Introduction}

It is not straightforward to give an invariant definition of mass in an asymptotically de Sitter space-time containing mass in a compact region, due to problems arising from the existence of a cosmological event horizon.
This is an old question that has been addressed by many authors, for example \cite{Abbott+Deser,Nakao, Shiromizu,CJJ,Dehghani,G+M} among others.
  An approximate definition, that works well provided any black-hole horizon $r_{BH}$ is very much smaller that than the cosmological horizon $r_C$, was given in \cite{Abbott+Deser}. In that work it is assumed that there is 
a region $r_{BH} << r <r_C$ in which the full de Sitter group $SO(1,4)$ is an approximate symmetry and a time-like generator is used to define the energy density, from which a mass is obtained by integrating over a 2-sphere of radius $r$.  The resulting mass is a good candidate provided corrections of order $\frac{r_{BH}}{r}$, but still with $r<r_C$, can be ignored.  Building on the work of Wald and collaborators \cite{Lee+Wald,Wald2,Iyer+Wald} we give in the following an exact definition of mass in asymptotically de Sitter space-times which only requires the existence of a time-like Killing vector at some $r$ with $r_{BH} < r < r_C$, but there is no approximation requiring $r>>r_{BH}$ and the full $SO(1,4)$ symmetry is not necessary --- one time-like Killing vector is sufficient.

The basic problem is most easily appreciated by examining a static asymptotically de Sitter Schwarzschild black hole with line element
\beq
d s^2 = -\left(1 - \frac{2 m}{r} -\frac{\Lambda r^2}{3} \right) d t^2
+\left(1 - \frac{2 m}{r} -\frac{\Lambda r^2}{3} \right)^{-1} d r^2
+ r^2(d\vartheta^2 + \sin^2\vartheta d\varphi^2)\label{SdeS}
\eeq
for which the Killing vector $\frac{\partial}{\partial t}$ is space-like for $r>r_C$,
where $r_C$ is the cosmological horizon --- the largest root of
the cubic equation
\beq
\Lambda r^3 - 3 r + 6 m  =0.
\eeq
The most widely accepted definition of mass in general relativity
involves identifying an asymptotically time-like Killing vector at either spatial infinity \cite{ADM,Brown+York} or light-like infinity \cite{Bondi}
and evaluating the energy over a 2-sphere there,  but there is no such time-like Killing vector for an asymptotically de Sitter black hole. 
This is particular vexing as the cosmological constant is measured to be positive \cite{LambdaPositive} so in principle we do not have a rigorous definition of mass for a black hole in our Universe (though the observed $\Lambda$ is so small that the definition in \cite{Abbott+Deser} should suffice for all practical purposes
for any known astrophysical black hole).  Nevertheless it would be gratifying to have a more mathematically rigorous definition.

In this work we give an invariant definition of mass for a black hole in de Sitter space-time that does not rely on taking $r\rightarrow\infty$, all that is necessary is that at some value of $r<r_C$ there is a time-like Killing vector in a region of space-time containing a 2-sphere surrounding the mass. The definition is
in essence very like Gauss' law in electrostatics, though in detail it is a lot more involved.  The bottom line is that the mass can be calculated by integrating over a sphere of any radius as long as it completely surrounds the mass and is in a region with a time-like Killing vector. We give the example of Scwharzschild-de Sitter space-time, where the calculation can be done analytically for any radius $r$ and explicitly shown to be independent of $r$. In an appendix we also treat the Kerr-de Sitter metric where the calculation can only be done analytically as $r\rightarrow \infty$, but the general formalism ensures that the same answer would be obtained for finite $r<r_C$ were it possible to push it through analytically. The formalism promises to have applications in gravitational wave physics as metric perturbations 
corresponding to a gravitational wave, $\delta g$, can depend on time in a region where the background metric has a time-like Killing vector.  For realistic values of the cosmological constant numerical calculations could be performed in a region where $r$ is large enough for $\frac{\delta g}{r^2}$ to be small but still $r<r_C$.

The construction relies on the work of Wald and collaborators \cite{Lee+Wald,Wald2,Iyer+Wald} in which a Noether form associated with a Killing vector was identified which can be used to give a Noether charge associated to the symmetry
generated by the Killing vector. 
It was shown in \cite{Hajian+Jabbari} that Wald's formalism leads to the canonically accepted 
Henneaux-Teitelboim mass for the asymptotically anti-de Sitter Kerr metric and the analysis in the appendix can be succinctly summarized in the statement that the Wald mass of the asymptotically de Sitter Kerr metric is simply the analytic continuation of \cite{Hajian+Jabbari} from negative to positive $\Lambda$.

\section{The invariant mass}

In preparation for the main calculation of the paper we first summarize the construction of Lee and Wald's invariant mass \cite{Lee+Wald,Wald2}
(a fuller treatment using the language of differential forms is given in \cite{BPD1}).
Consider a theory with fields $F^I$ governed by a Lagrangian
density $L(F^I,\partial_\mu F^I)$ (a $D$-form) on a $D$-dimensional
space-time (in this section $D$ is arbitrary, later we shall  specialize to $D=4$). Under a variation of the fields, $F^I\rightarrow F^I + \delta F^I$, the variation of $L$ yields the equations of motion,
$E_J(F^I)$, together with a total derivative,
\[ \delta L = E_J(F^I)\delta F^I + d\theta(F^I,\delta F^I)\]
where $\theta$ is a $(D-1)$-form.
Let ${\cal S}$ denote the (infinite-dimensional) space of all possible
solutions of the equations of motion.
For any specific solution of the equations of motion, $F^I=S^I\in {\cal S}$,
\[ \delta L =  d\theta(S^I,\delta S^I).\]
$\theta(S^i,\delta S^I)$ is linear in the infinitesimal variation $\delta S^I$ (which is a solution of the linearized equations of motion) and can be thought of as a 1-from on $T^*{\cal S}$, as well as a $(D-1)$-form on ${\cal M}$.  Under a second distinct field variation,
anti-symmetrised under the two variations, define
\beq \omega(S^I,\delta_1 S^I, \delta S^I_2) = \delta_2 \theta(S^I,\delta S^I_1)
- \delta_1 \theta(S^I,\delta S^I_2).\label{omegadef}\eeq
$\omega$ a 2-from on the space of solutions and we can write this (\ref{omegadef}) as
\[ \omega = \delta \theta\]
with $\delta$ being viewed as the exterior derivative on the space of solutions,
satisfying $\delta^2 = 0$ and $\delta d = d \delta$.

Given a foliation of space-time ${\cal M}$ into space-like hypersurfaces $\Sigma_t$, labelled by a monotonic time-parameter $t$,
\[ \Theta = \int_{\Sigma_t} \theta\]
is a (pre-)symplectic potential for a (pre-)symplectic structure
\[ \Omega =\int_{\Sigma_t}\omega = \delta \Theta\]
on the phase space of the theory. Since
\[ d\omega = d \delta \theta = \delta d \theta = \delta^2 L =0\]
on shell we can deduce that, with suitable conditions on the fall of the
fields at the boundary, $\partial \Sigma_t$, $\Omega$ is independent
of the choice of hypersurface $\Sigma_t$.

For a field theory invariant under diffeomorphisms and gauge transformations
we shall denote the group of all such transformations by ${\cal G}$ and denote the space of solutions of the equations of
motion mod gauge transformations and diffeomorphisms by $\widehat {\cal S}={\cal S}/{\cal G}$.
Under a projection from 
${\cal S}$ to ${\cal S}/{\cal G}$ there must be a symplectic structure
${\widehat \Omega}$ on $\widehat{\cal S}$ that pulls back to $\Omega$ on
${\cal S}$ under the projection. For this to be true $\Omega$ should
vanish when one of the field variations is a diffeomorphism, generated
by a vector field $\vec X$ say. This will be the case if the symplectic density is $d$-exact when one  of the variations is a diffeomorphism, 
 \[\omega(\vec X) = d \phi(\vec X), \]
for some $(d-2)$-form $\phi (\vec X)$ depending linearly on $\vec X$.
Since then, assuming that $\vec X$ vanishes on (and falls of sufficiently fast 
near) the boundary $\partial \Sigma$ of $\Sigma$ \cite{C-W}, 
we have\footnote{Since $\phi(\vec X)$ is linear in $\vec X$ it vanishes on $\partial \Sigma$ if $\vec X$ vanishes there.}
\[ \Omega[\vec X] = \int_{\Sigma} d \phi(\vec X) = \int_{\partial \Sigma} \phi(\vec X) =0.\]

Under an infinitesimal  diffeomorphism, generated by a vector field $\vec X$,
the variation of the Lagrangian is
\[ \delta L  = {\cal L}_{\vec X}L =  d i_{\vec X} L\]
where ${\cal L}_{\vec X} = d i_{\vec X} + i_{\vec X} d $ denotes the action of
the Lie derivative on differential forms.  Thus on shell
\[ d\theta(\vec X) = d i_K L\qquad \Rightarrow \qquad \theta(\vec X) = i_{\vec X} \theta(\vec X) +  J(\vec X)\]
with $J(\vec X)$ is a closed $(D-1)$-form, the Noether current of \cite{Wald2}.
Assuming following \cite{Wald1} that, with reasonable assumptions, 
$J(\vec X)$ is exact, so $J(\vec X)=d Q(\vec X)$ for some $(D-2)$-form $Q(\vec X)$ depending linearly on the diffeomorphism $\vec X$, we can write
\beq\theta(\vec X) = i_{\vec X} L + d Q(\vec X). \label{thetaQ}\eeq
When one of the variations is a diffeomorphism we have
\beq\omega(\vec X) =  d \phi(\vec X)=\delta \theta(\vec X) - {\cal L}_{\vec X} \theta \label{omega_X}\eeq
where  $\theta$ in the last term depends on a variation which corresponds to a physical variation of the fields and is not just a diffeomorphism.
Since $\delta L = d \theta$ on shell and $i_{\vec X} \delta = \delta i_{\vec X}$ 
we have, using (\ref{thetaQ}) and 
(\ref{omega_X}),
\[\omega(\vec X) = i_{\vec X}\delta L  + d \delta Q(\vec X) - d i_{\vec X}\theta - i_{\vec X}d\theta
=  d (\delta Q(\vec X) - i_{\vec X}\theta). \]

If a Hamiltonian $h(\vec X)$ exists which generates the flow $\vec X$ then we expect
\[\omega(\vec X) = -\delta h(\vec X) =d(\delta Q(\vec X) - i_{\vec X}\theta). \]
Whether or not such a Hamiltonian does exist will depend on the theory, it is only possible if $i_{\vec X}\theta$ is $\delta$-exact, 
\[ i_{\vec X}\theta = \delta \mu(\vec X)\]
for some $(D-2)$-form  $\mu(\vec X)$ on ${\cal M}$ and this may not be the case. It is however true in
general relativity \cite{Lee+Wald} and we shall assume that such a $\mu(\vec X)$ exists.

 Furthermore if $\vec X=\vec K$ is Killing then $\omega(\vec K)$ 
vanishes identically \cite{Lee+Wald}
\[ \delta h(\vec K) =d\delta\bigl( \mu(\vec K) - Q(\vec K) \bigr) =0\]
and the $(D-2)$-form 
\[\phi(\vec K)=\delta \bigl(Q(\vec K) - \mu(\vec K)\bigr)\] 
is not only $\delta$-exact it is also $d$-closed.

In contrast to a general diffeomorphism we do not assume  that $\vec K$ vanishes on $\partial\Sigma$ so $\phi(\vec K)$ can also be non-zero there.  In that case
\[\delta \int_\Sigma h(\vec K) = -\int_{\Sigma}d \phi(\vec K)=
\delta \int_{\partial \Sigma}\bigl(\mu(\vec K) - Q(\vec K) \bigr)=0. \]
We shall refer to the $(D-2)$-form
\[\rho(\vec K) = \mu(\vec K) - Q(\vec K) \]
as the Noether form of the second kind (to distinguish it from $Q(\vec K)$ which is referred to as the Noether form in \cite{Wald2}).
Thus $\int_{\partial \Sigma}\rho(\vec K)$ is a function on the space of solutions which depends on ${\vec K}$
 with $\delta\int_{\partial \Sigma} \rho(\vec K)=0$.

For example suppose there is a mass in some compact region surrounded by a sphere of radius $r_0$ and
 $\partial \Sigma$ consists of two nested $(D-2)$-spheres, one inside the other, at radii
$r_0$ and $r_1>r_0$ (we may take $r_1\rightarrow \infty$, but this is not necessary). 
Then the boundary of $\Sigma$ consists of two pieces, a $(D-2)$-sphere at 
$r_0$ and another at $r_1$, $\partial \Sigma = S^{(D-2)}_{r_0} \cup S^{(D-2)}_{r_1}$.
We can define
\[ \delta  \int_{S^{D-2}|_{r_1}} \rho(\vec K)=\delta \int_{S^{D-2}|_{r_0}} \rho(\vec K) 
:=\delta {\cal Q}[\vec K]\]
and the variation $\delta {\cal Q}(\vec K)$ is independent of the value of $r$ at which it is calculated.
Thus 
\[{\cal Q}[\vec K]=\int_{S^{D-2}} \rho(\vec K) \]
is a candidate for a conserved quantity associated with the Killing vector $\vec K$.  

It is emphasized that $\delta {\cal Q}[\cal K]$ is independent of the value of
$r$ at which it is calculated and there is no requirement that the metric perturbation share the Killing symmetry of the solution, all that is required is that
it satisfy the linearized Einstein equations. 

\subsection{Invariant mass in general relativity}

For a 4-dimensional space-time ${\cal M}$ ($D=4$ in this section) with metric $g_{\mu\nu}$ and co-ordinates $x^\mu$ we foliate ${\cal M}$ with constant time hypersurfaces and let $x^\mu = (t,x^\alpha)$  where $\alpha=1,2,3$ and $t$ is a time co-ordinate.  
We use the standard ADM decomposition: $t=const$ are space-like hypersurfaces, $\Sigma_t$, and we denote the induced metric on $\Sigma_t$ by $h_{\alpha\beta}(t)$. The 4-dimensional line element decomposes as
\bea d s^2 &=& g_{\mu\nu} d x^\mu d x^\nu = g_{t t}^2 + 2 g_{t\alpha} d t d x^\alpha + g_{\alpha\beta} dx^\alpha d x^\beta\nonumber \\
&=&-N^2 d t^2 + h_{\alpha\beta} (d x^\alpha + N^\alpha d t)(dx^\beta + N^\beta dt),\eea
where $g_{t t}=-N^2 + h_{\alpha\beta} N^\alpha N^\beta$,
$g_{t\alpha}=g_{\alpha\beta} N^\beta $ and $ h_{\alpha\beta} = g_{\alpha\beta}$. 

We shall employ differential form  notation using
orthonormal 1-forms $e^a$ for the metric $g$, which can be expressed in a co-ordinate basis as
\[ e^a=e^a{}_\mu d x^\mu.\]
The connection 1-forms are determined by the torsion-free condition
\[D e^a = d e^a + \omega^a{}_b \wedge e^b=0 \]
and the curvature 2-forms are
\[R_{ a b} = \frac{1}{2} R_{a b c d} e^c \wedge e^c = d\omega_{a b} + \omega_a{}^c \wedge \omega_{c b}\]
where orthonormal indices are raised and lowered using $\eta^{a b} = \eta_{a b}
=\hbox{diag}(-1,+1,+1,+1)$.

Under the above foliation denote orthonormal 1-forms for $h_{\alpha \beta}$ by
\[\tilde e^i=\tilde e^i{}_\alpha dx^\alpha,\]
with $i=1,2,3$, so $\tilde e^i{}_\alpha = e^i{}_\alpha$ with
\beq  e^0=N dt \qquad\hbox{and}\qquad e^i=\tilde e^i + \frac{N^i}{N}  e^0,
\label{app:eetilde}\eeq
and $N^i=e^i{}_\alpha N^\alpha$ the orthonormal components of the shift vector.
The connection 1-forms associated with $\tilde e^i{}_\alpha$ on $\Sigma_t$ are defined using the zero torsion condition
\beq\tilde d\tilde e^i + \widetilde \omega^i{}_j \wedge \tilde e^j =0\label{omega-tilde-def}\eeq
with $\tilde d= \tilde e^i \partial_i$ the exterior derivative on $\Sigma_t$
at constant $t$.

In this gauge
\bea e^a{}_\mu = \begin{pmatrix}
N & 0 \\ N^i & \tilde e^i{}_\beta
\end{pmatrix}, \qquad
(e^{-1})^\mu{}_a = 
{\renewcommand*{\arraystretch}{1.2}
\begin{pmatrix}
\frac 1 N & 0 \\ 
-\frac{N^\alpha}{N} & (\tilde e^{-1})^\alpha{}_j
\end{pmatrix}}\label{eq:Vierbeins}
\eea
and the unit vector normal to $\Sigma_t$, $\vec n$, has orthonormal components $n^a=(1,0,0,0)$ so the metric dual 1-form is
$n=n_a e^a = -e^0$.

Metric variations are described by variations in the tetrad $\delta e^a = \delta e^a{}_\mu d x^\mu$
which can be encoded into the square matrix 
\beq 
{\Delta}^a{}_b  = ({\delta}  e^a{}_\mu)( e^{-1})^\mu{}_b =
{\renewcommand*{\arraystretch}{1.2}
\begin{pmatrix}
  \frac {{\delta} N} N & 0 \\  
 \frac{ \tilde e^i{}_\alpha {\delta} N^\alpha}{N} &  {\Delta}^i{}_j
\end{pmatrix}},\label{deltae}
\eeq
with $ {\Delta}^i{}_j = ({\delta} \tilde e^i{}_\alpha)(\tilde e^{-1})^\alpha{}_j$.  $\Delta_{i j}$ can be decomposed into symmetric and anti-symmetric parts
\[\ws_{i j} = \Delta_{\{i j\}}=\frac 1 2 ( {\Delta}_{i j} +  {\Delta}_{j i}), \qquad  {A}_{i j} = \Delta_{[i j]} = \frac 1 2 ( {\Delta}_{i j} -  {\Delta}_{j i}).\]

Let $S=S^i{}_i$ be the trace of $S_{i j}$ and $\kappa_{i j}$ be the extrinsic curvature\footnote{In the gauge (\ref{eq:Vierbeins})
$\kappa_{i j} = \frac 1 2 (D_i n_j + D_j n_i)$.} of $\Sigma_t$ and $\kappa=\kappa^i{}_i$ its trace.  

For the Einstein action with a cosmological constant,
\[ S = \int_{\cal M} (R_{a b} \wedge *(e^a \wedge e^b) - 2 \Lambda *1),\]
it was shown in \cite{BPD2} that,
if $\vec {K} = \frac{\partial}{\partial t}$ is Killing and 
$\Sigma_t$ can be foliated into 2-dimensional spheres $S^2|_{t,r}$ parameterized by $r$, then
\bea
\delta {\cal Q}[\vec K]
&=& \frac{1}{8\pi }\int_{S^2|_{t,r}}
\big\{
N\bigl(\wD_j\ws_i{}^j - \partial_i \ws)
+(\partial_i N) \ws - (\partial_j N)\ws_i{}^j\nonumber\\
&&\kern 50 pt
+{X}_{i j} N^j - N_i (\kappa_{j k} \ws^{j k} + {\delta}\kappa)
\big\}* e^{0 i},\label{delta-Q-def}
\eea
where $\widetilde D_j$ is the co-variant derivative associated with the orthonormal 1-forms $\tilde e^i$ (here
 \beq
{X}_{i j}={\delta} \kappa_{i j} + [\kappa,{\Delta}]_{i j}
+\kappa_{i j} \ws
\eeq
and $[\kappa,{\Delta}]_{i j}$ is the commutator of the matrices $\kappa_{i j}$ and ${\Delta}_{i j}$).
$\delta {\cal Q}[\vec K]$ is guaranteed to be independent of $t$ and $r$ and
 ${\cal Q}[\vec K]$ is the mass contained within ${S^2}|_r$.
For asymptotically flat space-times it corresponds to the ADM mass when $r\rightarrow \infty$ with $t$ fixed and ${S^2}|_r$ is space-like \cite{Iyer+Wald}, and it gives the Bondi mass when $r\rightarrow \infty$ with $(t-r)$ fixed and ${S^2}|_r$ is a null surface \cite{BPD2}.

A crucial observation is that it is not necessary to take the asymptotic limit as long as the perturbation satisfies Einstein's equations and the 2-surface ${S^2}|_r$  lies in a region where $\vec K$ is Killing \cite{Hajian+Jabbari,BPD2} --- it is not even necessary that the perturbation has the Killing symmetry. 

\subsection{Schwarzschild-de Sitter space-time}

As an example consider the Schwarzschild-de Sitter metric with line element 
(\ref{SdeS}) with Killing vector $\vec K=\frac{\partial}{\partial t}$.
We can choose orthonormal 1-forms
\[ e^0 = \sqrt{1-\frac{2 m}{r} - \frac{\Lambda r^2 }{3}}\,d t,
\qquad e^1 = \frac{1}{\sqrt{1-\frac{2 m}{r} - \frac{\Lambda r^2 }{3}}}d r,
\qquad e^2=r d \vartheta, \qquad e^3 = r \sin\vartheta d \phi\]
for $r$ in the region $r_{BH}<r<r_C$. Thus
\[ N= \sqrt{1-\frac{2 m}{r} - \frac{\Lambda r^2 }{3}}, \qquad
N^i=0.\]
$\kappa_{i j}=0$ and (\ref{delta-Q-def}) simplifies to 
\beq
\delta {\cal Q}[\vec K]
= \frac{1}{8\pi }\int_{S^2}
\big\{
N\bigl(\wD_j\ws_1{}^j - \partial_1 \ws)
+(\partial_1 N) \ws - (\partial_j N)\ws_1{}^j\bigr\} * e^{0 1}.\label{eq:delta-QK}
\eeq

Now
\[\partial_1 =  N\partial_r,\qquad \partial_2 = \frac 1 r \partial_\vartheta \qquad \hbox{and} \qquad 
\partial_3 = \frac 1 {r\sin\vartheta} \partial_\varphi \]
so $\partial_2 N = \partial_3 N=0$ and
\[\partial_1 N = \frac{m}{r^2} - \frac{\Lambda r}{3}.\]
The non-zero connection 1-forms arising from (\ref{omega-tilde-def}) are
\[\widetilde \omega_{12}=-\frac{N}{r} e^2,\qquad 
\widetilde \omega_{13}=-\frac{N}{r} e^3,\qquad 
\widetilde \omega_{2 3} = -\frac{\cot\vartheta}{r} e^3\]
from which\footnote{The notation here is $\widetilde \omega_{i j}=\widetilde \omega_{i j|k} \tilde e^k$.}
\[\wD_j\ws_1{}^j=\partial_j S_1{}^j + \widetilde \omega_{1 k|j} S^{k j}
+ \widetilde \omega^j{}_{k|j} S_1{}^k
= N \left( S'_{1 1} +\frac{2}{r} S_{1 1}  -\frac{1}{r} S_\perp\right) 
+\frac{1}{r} \left(\partial_\vartheta +\cot\vartheta \right) S_{1 2}
+\frac{1}{r} \partial_\varphi S_{1 3}\]
where $'=\frac{\partial}{\partial r}$ and
\[ S_\perp = S^2{}_2 + S^3{}_3 \]
is the transverse trace of $S_{i j}$.

Using this in (\ref{eq:delta-QK}) gives
\bea
\delta {\cal Q}[\vec K] 
&=& \frac{1}{8\pi }\int_{S^2}
\left\{
\frac{2 N^2 S_{1 1}}{r} - N^2(\ws_\perp)'
+ \left(\frac{m}{r^2} - \frac{\Lambda r}{3} -\frac{N^2}{r}\right)   
\ws_\perp \right.\\
& & \hskip 80pt \left.+\frac{N}{r}\left( (\partial_\vartheta +\cot\vartheta ) S_{1 2}
+ {1\over\sin\vartheta}\partial_\varphi S_{1 3}\right) \right\}
 r^2 \sin \vartheta d \vartheta d \varphi.\nonumber
\eea
Now $ \frac{1}{r}\bigl((\partial_\vartheta +\cot\vartheta ) S_{1 2}
+ {1\over \sin\vartheta}\partial_\varphi S_{1 3}\bigr)=\nabla_{\hat\imath} V^{\hat\imath}$ is the divergence of a vector field $V^{\hat \imath} = S_1{}^{\hat \imath}$ (with $\hat\imath = 2,3$)
generating a diffeomorphism of $S^2$
 and as such its integral over $S^2$ must
vanish from Stokes' theorem if $V^{\widehat\imath}$ is globally well defined on $S^2$. We finally arrive at 
\beq
\delta {\cal Q}[\vec K] 
= \frac{1}{8\pi }\int_{S^2}
\left\{
\frac{2 N^2 S_{1 1}}{r} - N^2(\ws_\perp)'
+ \left(\frac{m}{r^2} - \frac{\Lambda r}{3} -\frac{N^2}{r}\right)   
\ws_\perp \right\} r^2 \sin \vartheta d \vartheta d \varphi.
\eeq

In this expression we must remember that $S_{1 1}$ and $S_\perp$ are not independent, they are related by the condition that the metric variation must satisfy the linearized equations of motion with $\Lambda$ fixed, 
in this case  $\delta e^a$
must be such that the variation of the Ricci tensor $\delta {\cal R}_{a b}=0$
in an orthonormal basis.
This is satisfied for example suppose by choosing $S_\perp=0$ and 
demanding that $S_{1 1}$ arises solely from varying $m$ in the function $N$, $m\rightarrow m + \delta m$ with $\delta m$ constant. 
With $e^1=\frac{d r}{N}$,
${\delta e^1} = -\frac{\delta N}{N}e^1$ this produces 
\beq S_{1 1} =-\frac{\delta N}{N}=\frac{\delta m}{N^2 r}\label{deltaNoverN}\eeq 
so
\beq
\delta {\cal Q}[\vec K] 
= \frac{1}{8\pi }\int_{S^2}
\left(\frac{2\delta m}{r^2}\right) r^2 \sin \vartheta d \vartheta d \varphi
= \delta m \label{delta-Q-m}
\eeq
and indeed the parameter 
\[ m={\cal Q}[\vec K]\] 
is the Noether charge associated with the Killing vector $\vec K= \frac{\partial}{\partial t}$. Like Gauss' law in electrostatics the expression (\ref{delta-Q-m}) is valid for
any $r>0$, it is not necessary to take $r\rightarrow \infty$.  We are free to use any
value of $r>0$ to evaluate the mass analytically,
though only in the range $r_{BH}<r<r_C$ is $\frac{\partial}{\partial t}$
time-like.  Note that we cannot take $\delta N$ in (\ref{deltaNoverN}) to be a more general function of $r$ without allowing for $S_\perp \ne 0$ as $S_\perp$ must be determined by the linearized equations of motion.

For the Kerr-de Sitter space-time, with rotational parameter $a$ \cite{Carter}, the formalism gives
\beq{\cal Q}[\vec K]= \frac{m}{\left(1+\frac{a^2}{L^2}\right)^2}.
\label{M-Kerr}\eeq
The mathematical analysis is more involved in this case and is relegated to an appendix,  (\ref{Kerr-de-Sitter-mass}).  Indeed the calculation can only be pushed through analytically for 
$r\rightarrow \infty$, but the formalism guarantees that the final result for 
$\delta {\cal Q}[\vec K]$ is independent of $r$, and this could be checked numerically.  The result (\ref{M-Kerr}) is simply what one would obtain by analytically continuing the Henneaux-Teitelboim mass for the Kerr-anti-de Sitter space-time 
\cite{Hajian+Jabbari,Henneaux-Teitelboim} from negative to positive $\Lambda$.

\section{Conclusions}

For a large class of diffeomorphism invariant theories the formulation of Wald {\it et al.} allows an invariant charge to be associated with any solution of the equations of motion that admits a Killing vector. The Killing vector does not need to be globally defined, it suffices for it to be Killing outside of a compact region contained within a 2-sphere over which the charge is calculated, in a manner similar in spirit but different in detail to Gauss' law in electrostatics. The charge is calculated by perturbing the metric by a variation that satisfies the linearized equations of motion but need not share the Killing symmetry,
so the method is ideally suited to calculating gravitational mass with a perturbation corresponding to a gravitational wave provided a region can be isolated 
where the background metric is stationary inside the cosmological horizon.

Explicit examples have been given of Schwarzschild-de Sitter space-time, where
the calculation can be performed analytically for any value of $r$ and 
shown to be independent of $r$, and of Kerr-de Sitter space-time.
In the latter case the calculation cannot be done analytically at finite $r$ and is only pushed through for $r\rightarrow\infty$, but the general formalism ensures that the same value of the mass would be obtained for any value of $r$, in particular for $r<r_C$, though an explicit verification of this would require numerical calculation.  For a more general asymptotically de Sitter metric with a time-like Killing vector outside of some compact region one could use numerical computation to determine the mass if an analytic evaluation is not feasible.
  
 An important aspect of the formulation is that the charge can be calculated exactly by integrating over any sphere in a region of space where the Killing symmetry holds, it is not necessary to go to an asymptotic region.  For a time-like Killing vector this allows masses to be calculated in asymptotic de Sitter space-times, provided there is a region inside the cosmological horizon where the Killing symmetry holds.

The author acknowledges support by the Action
MP1405 QSPACE from the European Cooperation in
Science and Technology (COST) and wishes to thank the Macquarie University Research Centre in Quantum Science and Technology (QSCITECH), where part of this work was
carried out. for a Visiting Honorary Fellowship.

\appendix

\section{Asymptotically de Sitter stationary black holes \label{AsymptoticallydS}}

The line element outside a rotating black hole in de Sitter space-time
is \cite{Carter}
\beq  \label{dSKerr}
d s^2=-\frac{\Delta}{\Arho^2}\left( d \At - \frac{a\sin^2\Atheta}{ \Xi}\, d\Aphi \right)^2
+\Arho^2 \left( \frac{d\Ar^2} \Delta + \frac{d \Atheta^2} { \Xi_\Atheta} \right) 
+\frac{ \Xi_\Atheta \sin^2\Atheta}{\Arho^2} \left(\frac{\Ar^2 + a^2}{ \Xi}\,d\Aphi - a d \At  \right)^2
\eeq
with
\bea \label{MetricFunctions}
\Delta&=&\frac{(\Ar^2+a^2)(L^2 - \Ar^2)}{L^2} -2 m \Ar, \qquad
 \Xi_\Atheta=1+\frac{a^2}{L^2}\cos^2\Atheta,\nonumber \\
 \Arho^2 & = & \Ar^2 + a^2 \cos^2 \Atheta\qquad \hbox{and}  \qquad  \Xi = 1+\frac{a^2} {L^2}.
\eea
This can be decomposed into a pure de Sitter part and a part that vanishes when $m=0$,
\[ d s^2 = d s^2_{dS} + d s_m^2 \]
where
\bea d s_{dS}^2 &=& -\left(1-\frac{\Ar^2 + a^2\sin^2\Atheta}{L^2}  \right)d\At^2  
- 2 a \left(\frac{\Ar^2+a^2}{L^2} \right)\frac{\sin^2
  \Atheta}{ \Xi} d\At d\Aphi \nonumber \\
&&+\frac{ L^2 \Arho^2}{(\Ar^2 + a^2)(L^2-\Ar^2 )} d\Ar^2 
+\frac{\Arho^2 d\Atheta^2}{ \Xi_\Atheta}+(\Ar^2 + a^2)\frac{\sin^2\Atheta}{ \Xi}d\Aphi^2,\label{tildedSmetric}\\
d s_m^2 &=&
 \frac{2 m \Ar}{\Arho^2} \left(  d\At^2 - a \frac{\sin^2\Atheta}{ \Xi}  d\At d\Aphi  +\frac{\sin^4\Atheta}{ \Xi^2}d\Aphi^2 \right)\nonumber \\
&&+  \frac{ 2 m \Ar \Arho^2 L^4} {(\Ar^2+a^2)(L^2-\Ar^2)\bigl[(\Ar^2+a^2)(L^2-\Ar^2)+2 m \Ar L^2 \bigr]} d\Ar^2.
\eea
Despite appearances (\ref{tildedSmetric}) is the dS Sitter metric, but not in standard co-ordinates.
The co-ordinate transformation that puts it into a more standard form was given in \cite{Henneaux-Teitelboim}: with
\bean t&=&\At,\\
\varphi&=&\Aphi-\frac{a\At}{L^2},\\
r^2\cos^2\vartheta &=& \Ar^2\cos^2\Atheta,\\
r^2\sin^2\vartheta &=& (\Ar^2+a^2)\frac{\sin^2\Atheta}{ \Xi}.
\eean
Some useful relations are
\bea (L^2-\Ar^2  )\, \Xi_\Atheta &=&  (L^2-r^2 )  \Xi\\
\left(\frac{\Ar^2 + a^2}{\Ar^2}\right)\tan^2\Atheta &=&  \Xi \tan^2\vartheta.
\eea
One finds that (\ref{tildedSmetric}) is the more familiar
\[
d s^2_{dS}=-\left( 1-\frac{r^2}{L^2}  \right) d t^2 + \frac{1} {\bigl(1-\frac{r^2}{L^2}  \bigr)} d r^2 + r^2 (d\vartheta^2 + \sin^2\vartheta d \varphi^2).
\]

It is not illuminating to write $d s_m^2$ in $(t,r,\vartheta,\varphi)$ co-ordinates in general but we shall need its asymptotic form for $\Ar >>L$ (and $r>>L$).
Let
\[ \Sigma^2_\vartheta = 1 + \frac{a^2}{L^2} \sin^2 \vartheta\]
then some useful formulae for the deriving the asymptotic form of $d s^2_m$ in $(t,r,\vartheta,\varphi)$ co-ordinates are
\bean
 \Xi_\Atheta &=&  \frac{(L^2-r^2 )}{(L^2-\Ar^2  )}  \Xi =\frac{ \Xi}{\Sigma^2_\vartheta}+O\left( \frac{1}{r^2}\right)\\
\tan^2\Atheta &=&  \Xi \tan^2\vartheta+ O\left( \frac{1}{r^2}\right)\\
\Ar^2&=& \Sigma^2_\vartheta r^2 + O(1)\\
\cos^2\Atheta &=& \frac{\cos^2\vartheta}{\Sigma^2_\vartheta} + O\left( \frac{1}{r^2}\right)\\
\sin^2\Atheta &=&  \Xi \left(\frac{\sin^2\vartheta}{\Sigma^2_\vartheta}\right) + O\left( \frac{1}{r^2}\right).
\eean
Using these one finds
\[
\begin{pmatrix}
d \Ar \\ \Ar d \Atheta \\
\end{pmatrix} 
= 
\begin{pmatrix}
\Sigma_\vartheta & \frac{a^2 \cos\vartheta\sin\vartheta}{L^2 \Sigma_\vartheta} \\
0 &  \frac{\sqrt{ \Xi}}{\Sigma_\vartheta}\\
\end{pmatrix} 
\begin{pmatrix}
d r \\ r d \vartheta \\
\end{pmatrix} + O\left( \frac{1}{r^2}\right)
\]
and the leading terms in $d s_m^2$ are
\beq
d s^2_m= \frac{2 m}{r \Sigma^5_\vartheta} 
\big( d t - a \sin^2\vartheta d\varphi\bigr)^2
+\frac{2 L^4}{r^5 \Sigma^3_\vartheta} d r^2 + \cdots.
\eeq
In the time-gauge the vierbeins are of the form (\ref{eq:Vierbeins}) with leading order terms
\beq
e^a{}_\mu =
\begin{pmatrix}
\frac{r}{L}\sqrt{1-\frac{L^2}{r^2}} -\frac{m L}{r^2 \Sigma_\vartheta^5} & 0 & 0 & 0\\
0 & \frac{L}{r}\frac{1}{\sqrt{1-\frac{L^2}{r^2}}} +\frac{m L^3}{r^4 \Sigma_\vartheta^3} & 0 & 0 \\
0 & 0 & r +\frac{m a^2 L^2 \sin^2 (2\vartheta)}{r^4 \Sigma_\theta^6}  & 0 \\
-\frac{2 m a \sin\vartheta}{r^2\Sigma_\vartheta^5} & 0 & 0 & \left(1+ \frac{ma^2\sin^2\vartheta}{r^3 \Sigma_\vartheta^5}\right)r \sin\vartheta \\  
\end{pmatrix} + \cdots. \label{dS-Black-Hole-Vierbeins}
\eeq
In particular 
\[ N= \frac{r}{L}\sqrt{1-\frac{L^2}{r^2}} -\frac{m L}{r^2 \Sigma_\vartheta^5}  + O\left(\frac{1}{r^3}\right).\]
The $m/r^4$ term in $e^2{}_r$ is retained because we define $f(r,\vartheta)$ via
\[ e^1{}_r=\frac{1}{f} \]
with
\[f= \frac{r}{L}\sqrt{1 - \frac{L^2}{r^2} }
 -\frac{m L}{r^2 \Sigma_\vartheta^3} + O\left(\frac{1}{r^3}\right),\]
while the $m/r^4$ term in $e^2{}_\theta$ does not affect the subsequent analysis and can be discarded.

 To order $\frac{1}{r^3}$ the connection 1-forms are
\bean
\omega_{01} &=& -\frac{\partial_r N}{N} f e^0, \qquad
\omega_{02} = -\frac 1 r \frac{\partial_\vartheta N}{N} e^0, \qquad 
\omega_{03}=0,\\
\omega_{12} &=& -\frac 1 r \frac{\partial_\vartheta f}{f} e^1  - \frac{f}{r} e^2, \qquad
\omega_{23} = -\frac{\cot\vartheta}{r} e^3, \qquad \omega_{13} = -\frac{f}{r} e^3. 
\eean
Hence asymptotically
\bean N & \rightarrow & \frac{r}{L}, \qquad f  \rightarrow  \frac{r}{L},\\
\omega_{01,0} &\rightarrow& -\frac 1 L, \qquad  \omega_{12,2}\  \rightarrow \ -\frac 1 L, \qquad \omega_{13,3}\  \rightarrow \ -\frac 1 L.
\eean

We will now evaluate (\ref{delta-Q-def}) using this asymptotic
behaviour.
We have $\partial_1 = f \frac{\partial}{\partial r}+O\left(\frac{1}{r^2}\right) $  and, although $\partial_r N \sim 1/L$, for transverse derivatives
$\partial_{\hat \jmath} N \sim  O\left(\frac{1}{r^2}\right)$ and $\omega^j{}_{1,1} \sim O\left(\frac{1}{r^3}\right) $. One finds
\bean
\int_{S^2_{t,r}}  \Bigl[N \bigl\{ \widetilde D_j (\ws_i{}^j) - \partial_i (\ws_j{}^j)\bigr\} 
+\partial_i N \bigl( n^i \ws^j{}_j  - n^j \ws^i{}_j   \bigr)\Bigr]r^2 
\widehat e\,{}^{2 3} \\
=\int_{S^2} \left\{ N f \left (\frac{2}{r} \ws_{11} -\frac{1}{r} S_\perp  - (S_\perp)'\right)
  + N'f S_\perp  \right\}r^2 \widehat e\,{}^{23} +O\left(\frac 1 r\right)
\eean
where $S_\perp = \ws_{22} + \ws_{33}$ is the transverse trace of $\ws_{i j}$, $'=\frac{\partial}{\partial r}$ and 
\[\widehat e\,{}^{2 3}= \sin\vartheta d\vartheta \wedge d\varphi\] 
is the volume form on the unit sphere.
Now
\bean N f &=& -\frac{r^2}{L^2} + 1 -\frac m r \left(\frac{1}{\Sigma_\vartheta^5} + \frac{1}{\Sigma_\vartheta^3} \right) + O\left( \frac{1}{r^2}\right),\\
 N' f &=& -\frac{r}{L^2}   + O\left( \frac{1}{r^2}\right)\eean
so
\bean 
\int_{S^2} \left\{ N f \left ( \frac{2}{r} \ws_{11}  - \right. \right. && \left. \left. \kern -32pt \frac{1}{r} S_\perp  - (S_\perp)'\right)
  + N'f S_\perp  \right\}r^2 \widehat e\,{}^{23}
 \\
 &=&\int_{S^2} \left\{ \left(1-\frac{r^2}{L^2}\right)\left(2\frac{\ws_{11}}{r}
 - (S_\perp)'\right) - \frac {S_\perp} {r} 
\right\} r^2 \widehat e\,{}^{23} +O\left(\frac 1 r\right).
\eean

Now we can expand $\ws_{11}$ and $S_\perp$ in inverse powers of $r$ as
\bean S_\perp  &=& \sum_{n=1}^\infty \frac{ b_n(\vartheta,\varphi)}{r^n},\\
\ws_{11} &=& \sum_{n=1}^\infty \frac{ c_n(\vartheta,\varphi)}{r^n}
\eean
then
\bea 
\int_{S^2} &&\kern -20pt   \left\{ N f \left ( \frac{2}{r} \ws_{11}  -  \frac{1}{r} S_\perp  - (S_\perp)'\right)
  + N'f S_\perp  \right\} r^2 \widehat e\,{}^{23}
 \label{omega_ai_bi}\\
&=&-\int_{S^2}  \left\{\frac{ ( b_1+2  c_1)}{L^2} + \frac{2( b_2+ c_2)}{L^2 r} -  \frac{2 c_1}{r^2} + \frac{(3  b_3 + 2  c_3)}{L^2 r^2}\right\} r^2 \widehat e\,{}^{2 3}
+O\left(\frac 1 r\right).\nonumber
\eea

Averaging over the sphere let $ \bar b_i = \frac{1}{4\pi}\int_{S^2}  b_i(\vartheta,\varphi) \sin\vartheta d\vartheta d\varphi$ and $ \bar c_i = \frac{1}{4\pi}\int_{S^2}  c_i(\vartheta,\varphi) \sin\vartheta d\vartheta d\varphi$  
then,  for a finite expression in (\ref{omega_ai_bi}) as $r\rightarrow \infty$, we must demand that
\[  \bar b_1 + 2  \bar c_1 =0 \qquad \hbox{and} \qquad  \bar b_2 +  \bar c_2=0.\]
If we wish the deformation of the area of the sphere at infinity 
to remain finite we must further demand that $\bar b_1=0$, so $\bar c_1=0$
(if we wish the area to be invariant we impose the stronger restriction
$\bar b_1=\bar b_2=0 \Rightarrow \bar c_1=\bar c_2=0$).

In any case we finally arrive at 
\beq
\lim_{r\rightarrow\infty}\delta{\cal Q}[e^a, {{\cal L}}_{\vec {K}}e^a, \delta e^a]= 2\bar c_1
-\frac{1}{2 L^2}(3  \bar b_3 + 2  \bar c_3).\label{Omega_dS}
\eeq

For example if we vary the parameters $m\rightarrow m + \delta m$ and $a\rightarrow \delta a$ in the original metric (\ref{dS-Black-Hole-Vierbeins}),
keeping $L$ fixed, then $e^1=\frac{1}{f} dr$ and  
\[ \ws_{11} = -\frac{\delta f}{f} =  -\frac 1 2 \frac{\delta f^2}{f^2}=
\frac{L^2}{r^2} \delta\left(\frac{m}{r \Sigma_\vartheta^3} \right)+ \;O\left(\frac 1 {r^4} \right)=
 -\frac{L^2}{r^3}  \delta\left(\frac{m}{\Sigma_\vartheta^3}\right) + \;O\left(\frac 1 {r^4} \right),\]
with
\[ S_\perp  = \frac{1}{r\sin\vartheta}\delta  \left(\frac{m a^2 \sin^3\vartheta}{r^2 \Sigma_\vartheta^5}\right)+ \;O\left(\frac 1 {r^4} \right)
=  \frac{1}{r^3}  \delta\left(\frac{m a^2}{\Sigma_\vartheta^5}\right)+ \;O\left(\frac 1 {r^4} \right)\]
so
\[c_1=0,\qquad
 b_3 = \sin^2\vartheta \,
\delta\left(\frac{m a^2}{\Sigma_\vartheta^5}\right)  \qquad\hbox{and}\qquad  c_3 =- L^2 \delta \left(\frac{m}{\Sigma_\vartheta^3}\right).\]
(\ref{Omega_dS}) is therefore
\bean \lim_{r\rightarrow\infty}{\cal Q}[e^a, {{\cal L}}_{\vec {K}}e^a, \delta e^a] 
&=&\frac{1}{8\pi}\int_{S^2}\left\{ 2\,\delta\left(\frac{m}{\Sigma_\vartheta^3}\right) - \frac {3\sin^2\vartheta} {L^2}\, \delta\left(\frac{m a^2}{\Sigma_\vartheta^5}\right)\right\} \sin\vartheta d \vartheta d \varphi\\
&=&\frac{1}{8\pi} \delta\left\{\int_{S^2}m \left(\frac{2}{\Sigma_\vartheta^3} - \frac{3 a^2}{L^2}\frac{\sin^2\vartheta}{\Sigma_\vartheta^5}\right) \sin\vartheta d \vartheta d \varphi\right\}\\
&=& \delta \left\{ \frac{m}{4}\int_0^{2\pi}\left(\frac{2}{\Sigma_\vartheta^3} - \frac{3 a^2}{L^2}\frac{ \sin^2\vartheta
}{\Sigma_\vartheta^5}\right) \sin\vartheta d \vartheta \right\}.
\eean
The integrals are elementary,
\beq \int_0^{2\pi}\frac{\sin\vartheta d\vartheta }{\Sigma_\vartheta^3}=
\frac{2}{\bigl(1+\frac{a^2}{L^2}\bigr)},\qquad
\int_0^{2\pi}\frac{\sin^3\vartheta d\vartheta }{\Sigma_\vartheta^5}=
\frac{4}{3 \bigl(1+\frac{a^2}{L^2}\bigr)^2} ,\label{ElementaryIntegrals}\eeq
giving 
\[ \delta{\cal Q}[e^a, {{\cal L}}_{\vec {K}}e^a, \delta e^a] 
=  \delta M\]
with
\beq M=\frac{m}{\left(1+\frac{a^2}{L^2} \right)^2}.\label{Kerr-de-Sitter-mass}\eeq
This is actually the analytic continuation of the mass determined in \cite{Henneaux-Teitelboim}
(indeed it is presumably no co-incidence that the integrals (\ref{ElementaryIntegrals}) are precisely the ones
that appear in equation (B.7) of that reference).

Finally a note on normalization. For asymptotically flat space-time the normalization of the time-like killing vector $\vec K =\frac{\partial}{\partial t}$ is chosen that $\vec K$ has unit norm at $r\rightarrow \infty$.  This criterion cannot be used in asymptotically de Sitter space-time. For asymptotically de Sitter space-time the normalization can be fixed by using the natural normalization of the generators of the de Sitter group $SO(1,4)$ at $r\rightarrow \infty$, in which case the natural normalization is $\vec K= L\frac{\partial}{\partial t}$
and the invariant quantity obtained from this normalization is $m L$.


\begin{thebibliography}{7}

\bibitem{Abbott+Deser} L.F.~Abbott and S.~Deser, 
{\it Nucl. Phys. B} {\bf 195}, 76 (1982).
 
\bibitem{Nakao}  K.~Nakao, T.~Shiromizu and K.~Maeda, {\it Class. Quantum Grav.} {\bf 11}, 2059 (1994).

\bibitem{Shiromizu} T.~Shiromizu, {\it Phys. Rev. D} {\bf 60}, 064019 (1999), 
[arXiv:hep-th/9902049].

\bibitem{CJJ} P.T.~Chru\'sciel, J.~Jezierski and J.~Kijowski,
{\it Phys. Rev. D} {\bf 87}, 124015 (2013).

\bibitem{Dehghani} M.H.~Dehghani, 
{\it Phys. Rev. D} {\bf 65}, 104030 (2002), [hep-th/0201128].

\bibitem{G+M} A.M. Ghezelbash, R.B. Mann,
{\it Phys. Rev. D} {\bf 72}, 064024 (2005), [hep-th/0412300].

\bibitem{Lee+Wald} J.~Lee and R.M.~Wald, {\it J. Math. Phys.} {\bf 31}, 725 (1990).

\bibitem{Wald2} R.M.~Wald, {\it Phys. Rev. D} {\bf 48}, R3427 (1993), [arXiv:gr-qc/9307038].

\bibitem{Iyer+Wald} V.~Iyer and R.M.~Wald, {\it Phys. Rev. D} {\bf 50}, 846 (1994), [arXiv:gr-qc/9403028].


\bibitem{ADM} R.~Arnowitt, S.~Deser and C.W.~Misner,  {\it Phys. Rev.} {\bf 117}, 1595 (1960).

\bibitem{Brown+York} J.D.~Brown and J.W.~York, Jr, {\it Phys. Rev. D} {\bf 47}, 1407 (1993).

\bibitem{Bondi} H.~Bondi, 
{\it Nature} {\bf 186}, 535 (1960).

\bibitem{LambdaPositive}  N.~Aghanim {\it et al}, {\it A\&A}, {\bf A13}, 594 (2016), [arxiv:1502.1589].


\bibitem{Hajian+Jabbari} K.~Hajian and M.M.~Sheikh-Jabbari, {\it Phys. Rev. D}, {\bf 93} 044074 (2016), [arXiv:gr-qc/9307038].

\bibitem{BPD1} B.P.~Dolan, {\it Phys. Rev. D} {\bf 98}, 044009 (2018), [arXiv:1804.07689].

\bibitem{C-W} C.~Crnkovi\'c and E.~Witten, {\sl Covariant description of canonical formalism in geometrical theories} in {\it Three Hundred Years of Gravitation}, eds W.~Israel and S.W.~Hawking, 
CUP (1987).

\bibitem{Wald1} R.M.~Wald, {\it J. Math. Phys.} {\bf 31}, 2378 (1990).
 
\bibitem{BPD2} B.P.~Dolan, 
{\it Phys. Rev. D} {\bf 98}, 044010 (2018), [arXiv:1804.10451].

\bibitem{Carter} B.~Carter in {\it Les Astres Occlus} ed. by B. DeWitt, C. M. DeWitt, (Gordon and Breach, New York, 1973).

\bibitem{Henneaux-Teitelboim} M.~Henneaux and C.~Teitelboim, {\it Commun. Math. Phys.} {\bf 98}, 391 (1985). 

\end{thebibliography}
\end{document}